\documentclass[twocolumn]{revtex4-1}
\usepackage{graphicx}
\usepackage{epsfig}
\usepackage{amsmath}
\usepackage{amsfonts}
\usepackage{float}
\usepackage{hyperref}
\usepackage{amssymb}
\usepackage{color}
\usepackage[bottom]{footmisc}
\usepackage{array,multirow,makecell}

\def\lsim{\raise0.3ex\hbox{$\;<$\kern-0.75em\raise-1.1ex\hbox{$\sim\;$}}}
\def\gsim{\raise0.3ex\hbox{$\;>$\kern-0.75em\raise-1.1ex\hbox{$\sim\;$}}}
\def\be{\begin{equation}}
\def\ee{\end{equation}}
\def\bea{\begin{eqnarray}}
\def\eea{\end{eqnarray}}

\begin{document}

\title{A $U(1)_R$ inspired inflation model in no-scale Supergravity}
\author{ Shaaban Khalil$^{1}$, Ahmad Moursy$^{2}$, Abhijit Kumar Saha$^3$ and Arunansu Sil$^{3}$}
\affiliation{$^{1}$Center for Fundamental Physics, Zewail City of
Science and Technology, 6th October city,12566, Giza, Egypt.}
\affiliation{$^2$ Department of Mathematics, German University in Cairo, New Cairo City 11835, Egypt.}
\affiliation{$^3$ Department of Physics, Indian Institute of Technology Guwahati, 781039 Assam, India.}
\begin{abstract}
We consider a cosmological inflation scenario based on a no-scale supergravity sector with
$U(1)_R$ symmetry. It is shown that a tree level $U(1)_R$ symmetric superpotential alone
does not lead to a slowly rolling scalar potential. A deformation of this tree level superpotential
by including an explicit $R$ symmetry breaking term beyond the renormalizable level is proposed.
The resulting potential is found to be similar (but not exactly the same) to the one in Starobinsky
inflation model. We emphasize that for successful inflation, with the scalar spectral index $n_s
\sim 0.96$ and the tensor-to-scalar ratio $r < 0.08$, a correlation between the mass parameters
in the superpotential and the vacuum expectation value of the modulus field $T$ in the K\"ahler
potential must be adopted.

\end{abstract}
\maketitle
%
\section{Introduction}

Planck satellite's four years data of the cosmic microwave background radiation and the large structure
in the universe support the predictions of cosmological inflation. The recent data confirmed that
spectral index (scalar density fluctuations) is given by $n_s = 0.96 \pm 0.007$ and the upper bound
on the tensor-to-scalar ratio is  $r < 0.08$ \cite{Akrami:2018odb,Ade:2018gkx}. These results imposed
severe challenges on several inflationary models. For example, the simple chaotic and hybrid inflationary
models \cite{Dvali:1994ms} are now ruled out. On the other hand, some other models of inflation with
compatible cosmological fluctuation predictions receive a growing interest. One of these models is the
Starobinsky inflation \cite{Starobinsky:1980te}, which is based on modified gravity.

A supergravity (SUGRA) realization of Starobinsky inflation has been studied in Ref.\cite{Ellis:2013xoa}, by
considering a no-scale K\"ahler potential involving a modulus field $T$. It is well known that the no-scale
SUGRA framework is free of the so-called $\eta$ problem due to the involvement of logarithimic form in
the K\"ahler potential. In Ref.\cite{Ellis:2013xoa}, the no-scale K\"ahler potential of $T$ field is
combined with a Wess-Zumino superpotential consists of a quadratic and a cubic terms of the inflaton
superfield $S$: $W = \mu S^2 - \kappa S^3$, with $\mu$ as a parameter of mass dimension and
$\kappa$ is a dimensionless parameter. It turns out that at a specific point of the parameter space,
this construction becomes conformal equivalent to $R + \alpha R^2$ modified gravity models similar to
Starobinsky inflation model.  Adding a term linear in field $S$ to this renormalizable superpotential, the authors of
\cite{Romao:2017uwa} have shown that it is also possible to realize supersymmetry breaking at the
end of inflation. It is further indicated that a successful inflation consistent with correct $n_s$ and $r$
values may indicate an upper bound on gravitino mass, once the Starobinsky limit is implemented.
Few other studies having different kinds of motivation involving Starobinsky type
 inflation model can be found in \cite{Chakravarty:2014yda,Hamaguchi:2014mza,Pallis:2013yda,Kehagias:2013mya,Farakos:2013cqa,Copeland:2013vva,Garg:2015mra,Terada:2014uia,Garg:2017tds,Chakravarty:2017hcy,Ellis:2017jcp,Addazi:2017kbx,Ellis:2016ipm}.

While the constructions in \cite{Ellis:2013xoa} and \cite{Romao:2017uwa} are certainly elegant and
minimal from their own perspectives, we notice that it is not possible to define an $R$ charge for the
superfield $S$ so that superpotential $W$ can have $R$ charge of two units. Hence no $U(1)_R$
symmetry is prevailing in this construction. Now it is well known that $R$ symmetry plays important
roles in many supersymmetric constructions. One such example is related to the supersymmetry breaking.
According to Nelson-Seiberg theorem \cite{Nelson:1993nf}, existence of an $R$ symmetry is a
necessary condition in order to realize supersymmetry breaking. However an exact $R$ symmetry
forbids gauginos and Higgsinos to have mass. Hence it must be broken (spontaneously or explicitly).
It is customary to break $R$ symmetry spontaneously as done in many dynamical supersymmetry
breaking models leading to $R$ axions \cite{Bagger:1994hh}.

In this letter, we start with a $U(1)_R$ global symmetry. We assume that the inflaton superfield $S$
has an $R$ charge unity. Thus, the tree level superpotential is given by $W=\frac{\mu}{2}S^2$, with
$\mu$ as a mass scale. As we will show below, this tree level superpotential does not lead to a slowly
rolling scalar potential. We propose a deformation of this tree level
superpotential (having $R$ charge 2) by including an explicit $R$ symmetry breaking term beyond the
renormalizable level. This new term is naturally expected to be suppressed by the cut-off scale
$M_*$,
and hence $W$ can be expressed as
\begin{equation}
W=\frac{1}{2} \mu S^2 - \frac{1}{4}\lambda \frac{S^4}{M_*},
\label{superpotential}
\end{equation}
where $\lambda$ is a dimensionless coupling and $M_*$ is a mass scale.
Since a global symmetry is expected to be broken by gravity effects, a natural choice of $M_*$ would be
the Planck scale ($M_P$), $M_* \sim M_P$. A similar $M_P$ suppressed $R$-symmetry breaking term has been
considered in supersymmetric hybrid inflation scenario \cite{Civiletti:2013cra}, with minimal K\"ahler potential.
It was emphasized there that in order to get $n_s$ within the preferred range, $\lambda$ must
be small enough ($\lsim 10^{-7}$).
Below we study the above superpotential in Eq.(\ref{superpotential}) and discuss how inflation can be realized in this framework.

The paper is organized as follows. { In section II, we study the associated inflation model originated from
an interplay between a tree level $U(1)_R$  symmetric superpotential and a higher order explicit $R$-symmetry breaking
term along with no-scale K\"ahler potential.} In section III, inflationary predictions are discussed, in particular the correlation between the spectral index $n_s$ and the ratio $r$. Finally, our concluding remarks are given in section IV.

%
\section{The model} \label{sec:model1}
%
In addition to the superpotential $W$, we consider the K\"ahler potential  (as standard in no-scale supergravity)
\begin{align}
K=  -3 M_P^2 \ln \left[\frac{T+T^*}{M_P} - \frac{|S|^2}{3 M_P^2}\right],\label{Kahler}
\end{align}
where $T$ is the modulus field. The K\"ahler potential remains invariant under $U(1)_R$ symmetry with
vanishing $R$ charge for the moduli field.  The supergravity potential can be obtained using
\be
V_F=e^{\frac{K}{M_P^2}}\Big[(D_{j}W^*)(K^{-1})^j_i(D^iW)-\frac{3|W|^2}{M_P^2}\Big],
\ee
where,
\begin{align}
& D^iW=\frac{\partial W}{\partial \phi_i}+\frac{W}{M_P^2}\frac{\partial K}{\partial\phi_i}, & D_jW^*=\frac{\partial W^*}{\partial \phi^{*j}}+\frac{W^*}{M_P^2}\frac{\partial K}{\partial\phi^{*j}},\nonumber\\
 & K^j_i=\frac{\partial^2K}{\partial\phi^{*i}\partial\phi_j}, &,\nonumber
\end{align}
where $i,j$ refer to the modulus $T$ and inflaton $S$. Now using the superpotential in Eq.(\ref{superpotential}) and K\"ahler potential in Eq.(\ref{Kahler}),
$V_F$ can be obtained as
{
\begin{align}
V_F=\frac{1}{\Big(\frac{T+T^*}{M_P} -\frac{|S|^2}{3M_P^2}\Big)^2}\Big|\frac{\partial W}{\partial S}\Big|^2.
 \end{align}
 }
This is a feature of no-scale supergravity that
leaves the potential $V_F$ as independent of $T$ (apart from the dependence through the pre-factor $e^{K/M_P^2}$) and positive definite. Therefore, it can be an appropriate framework  for inflationary scenarios.

Following \cite{Ellis:2013xoa}, we assume here the modulus filed $T$ is stabilized at a fixed scale such that
$\langle T+T^* \rangle = c$. This stabilization requires a non-perturbative effect at a high
scale \cite{Ellis:2013nxa, Ellis:1984bs}. With this assumption, the effective Lagrangian turns out to be
\begin{align}
 \mathcal{L}_{\rm eff}= \mathcal{L}_{\textrm{K.E}}-V_F & = \frac{c/M_P}{\Big(\frac{c}{M_P}-\frac{|S|^2}{3M_P^2}\Big)^2}(\partial_\mu S^*)(\partial^\mu S) \nonumber\\
 &-\frac{1}{\Big(\frac{c}{M_P}-\frac{|S|^2}{3M_P^2}\Big)^2}\Big|\mu S-\lambda \frac{S^3}{M_*}\Big|^2\!.\label{PotSug}
\end{align}
In order to have the kinetic term for the complex scalar field $S$ as a canonically normalized one,
following the prescription of \cite{Ellis:2013xoa}, we first redefine the $S$ field in terms of $\chi$,
\begin{align}
S=\sqrt{3cM_P}\tanh\Big(\frac{\chi}{M_P\sqrt{3}}\Big). \label{refield}
\end{align}
With the above definition of $S$ and considering $\chi=(\chi_1+i\chi_2)/{\sqrt{2}}$,
the kinetic term ($\mathcal{L}_{\rm K.E.}$) becomes
\begin{align}
 \mathcal{L}_{\rm K.E.}=\sec^2\Big(\frac{2\chi_2}{\sqrt{3}M_P}\Big)(\partial_\mu\chi^*)(\partial^\mu\chi),
 \label{KE}
\end{align}
and the F-term scalar potential responsible for inflation
will have the form
\begin{align}
 V_F &=
  3M_P^4\Big(\frac{\mu^2}{M_P^2}\Big)\Big(\frac{M_P}{c}\Big) \Bigg[1- \Big| \tanh\Big(\frac{\chi_1+i \chi_2}{\sqrt{6}M_P}\Big) \Big|^2\Bigg]^{-2}\times \nonumber \\ &
   \Big|\tanh\Big(\frac{\chi_1+i \chi_2}{\sqrt{6}M_P}\Big)-
3\frac{c\lambda M_P}{\mu M_*}\tanh^3\Big(\frac{\chi_1+i \chi_2}{\sqrt{6}{M_P}}\Big)\Big|^2.
\label{Eq:VFF}
\end{align}
From this potential, one can show that the field dependent mass squared of the imaginary component of
$\chi$, obtained by the second derivative of $V_F$ respect to $\chi_2$ at the minimum $\chi_2=0$, is
much larger than the Hubble scale squared during inflation (a numerical estimate will be provided in next
section). Therefore  imaginary part $\chi_2$ will be stabilized at zero during the inflation.
Hence we set $\chi_2$ to be zero from now on and identify the F-term potential with the inflation potential,
$V_{\rm Inf}$. Note that with this choice, the kinetic term of the Lagrangian (see Eq.(\ref{KE})) becomes
canonical.

\begin{figure}[H]
\includegraphics[height=6 cm, width=8 cm,angle=0]{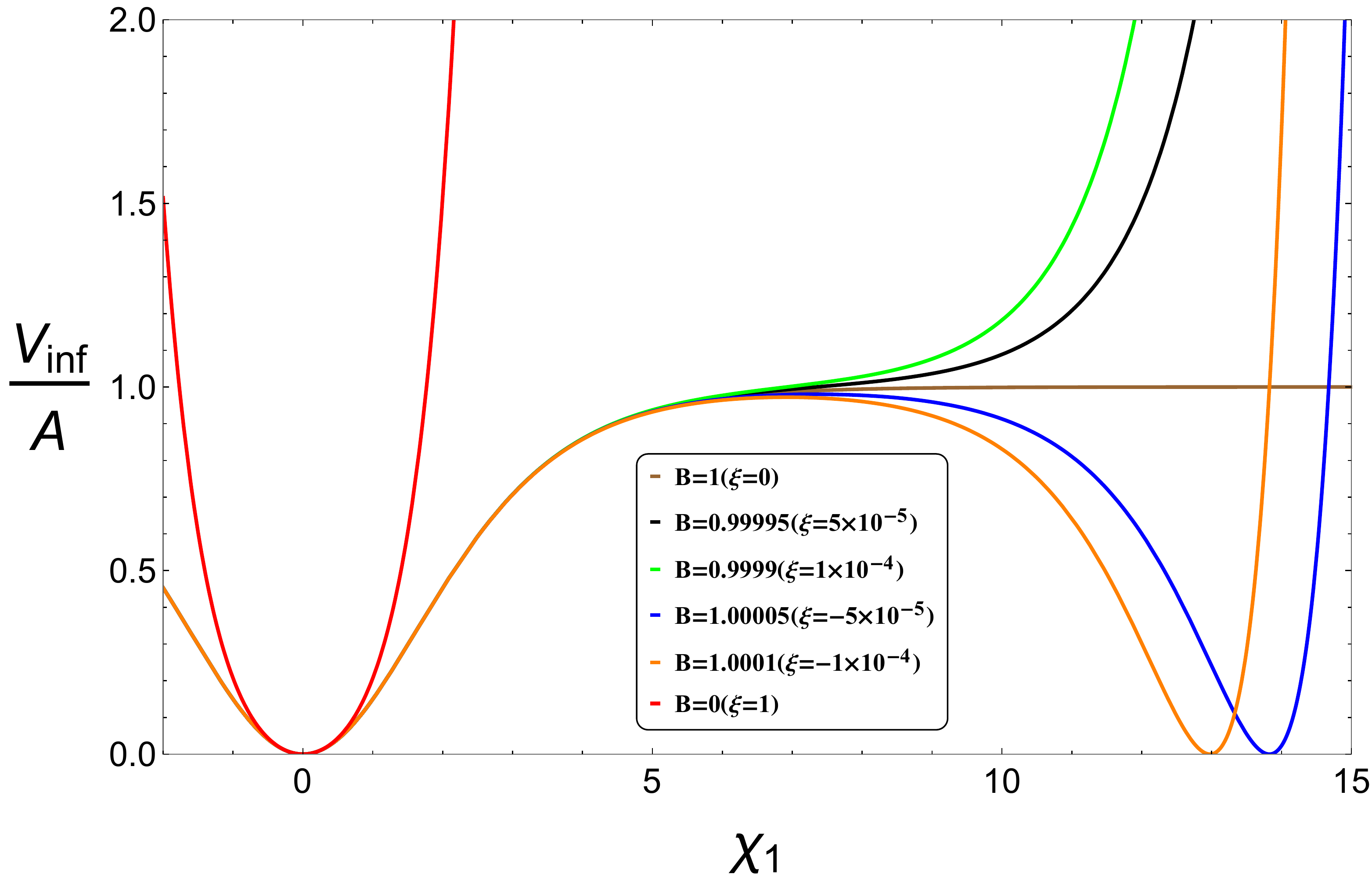}\\
\caption{Schematic plots of the potential $V_{\textrm{Inf}}/A$ against $\chi_1$ for different values of $B$ ($\xi$) around $B$=1 $(\xi =0)$.}
\label{MEig}
\end{figure}

In this case, the inflation potential takes the form,
\begin{align}
 V_{\textrm {Inf}} =
 A  \cosh^4\Big(\frac{\chi_1}{\sqrt{6}}\Big)\,\tanh^2\Big(\frac{\chi_1}{\sqrt{6}}\Big) \Big[1-
B\tanh^2\Big(\frac{\chi_1}{\sqrt{6}}\Big)\Big]^2,\label{Eq:Vinf}
\end{align}
where $A=3\frac{\mu^2}{c M_P}$ and $B=\lambda  \frac{3c M_P}{\mu M_*}$ are two dimensionless constants.
In the last expression of $V_{\textrm {Inf}}$, we have set $M_P=1$ unit. Untill otherwise stated, we will use this unit
for the rest of our discussion. Note that when $B=1$, this potential simplifies to the form
\begin{align}
 V_{\textrm{Inf}}^{(B=1)}= A \tanh^2\Big(\frac{\chi_1}{\sqrt{6}}\Big) \label{Eq:VinfB1}.
\end{align}
We have shown the form of this potential (normalized by $A$) in Fig. \ref{MEig} for different choices of $B$. With $B=1$, the
shape of the $V_{\textrm{Inf}}^{(B=1)}$ (denoted by the brown line in Fig.\ref{MEig}) turns similar to
the standard Starobinsky potential \cite{Starobinsky:1980te}. If we reduce the value of $B$ from 1 by
a tiny amount, the potential starts to become steep. On the other hand, if we enhance $B$ from one,
another distant minimum appears (other than at $\chi_1=0$) at some very large value of
the field $\chi_1$.

The inference of the above discussion is that the field $\chi_1$ can now be identified as the inflaton in the
limit (i) $\chi_2 \rightarrow 0$ and (ii) $B$ is very close to 1 as the required flatness for inflation is obtainable
from the associated potential $V_{\textrm{Inf}}$ of Eq.(\ref{Eq:Vinf}).
In order to show the importance of the $R$ symmetry breaking term  $\lambda$,  we
include a plot of the potential against $\chi_1$ with $\lambda = 0$ ({\it{i.e}} $B = 0$) in Fig. \ref{MEig} denoted
by the red curve. It is evident that such a potential can not provide sufficient inflation. Hence inclusion of
an explicit $R$ symmetry breaking term becomes instrumental in realizing inflation and that too by a restricted amount.
From the nature of the plots,  it is expected that for any large
deviation of $B$ from unity, the slow roll of the inflaton might be spoiled.

In order to have a better control over different values of $B$, we parameterize the deviation of $B$ from 1
by $\xi$, $B = 1 - \xi$.  Then the inflation potential in Eq.(\ref{Eq:Vinf}) can be expanded for
small $\xi$ as
\begin{align}
V_{\textrm{Inf}}\simeq A \tanh(\frac{\chi_1}{\sqrt{6}})^2 \Big\{1+ 2 \xi  \sinh(\frac{\chi_1}{\sqrt{6}})^2\Big\}.
\label{InfPA}
\end{align}

\noindent In $M_P=1$ unit, the slow roll parameters are given by
\begin{align}
 \epsilon=\frac{1}{2}\Big(\frac{V^\prime_\textrm{Inf}}{V_\textrm{Inf}}\Big)^2,~~\eta=\frac{V_\textrm{Inf}^{\prime\prime}}{V_\textrm{Inf}}.
\end{align}
Number of e-folds is written as
\begin{align}
 N_e=\int_{x_e}^{x^*}\frac{1}{\sqrt{2\epsilon}} dx,
\end{align}
where $x_e$ is the inflaton field value at which inflation ends and $x^*$ corresponds to the crossing horizon value of the inflaton. The three inflationary observables:
tensor to scalar ratio ($r$), spectral index ($n_s$) and power spectrum ($P_s$) are provided by
\begin{align}
 &r\simeq 16\epsilon,\\
& n_s\simeq 1-6\epsilon+2\eta,\\
&P_s=\frac{V_{\textrm{Inf}}}{24\pi^2\epsilon}.
\end{align}
These observables are to be determined at $x=x^*$.

\section{Inflationary predictions}
%

\begin{figure}[h!]
\vspace{0.5cm}
 \centering
 \includegraphics[width=0.5\textwidth]{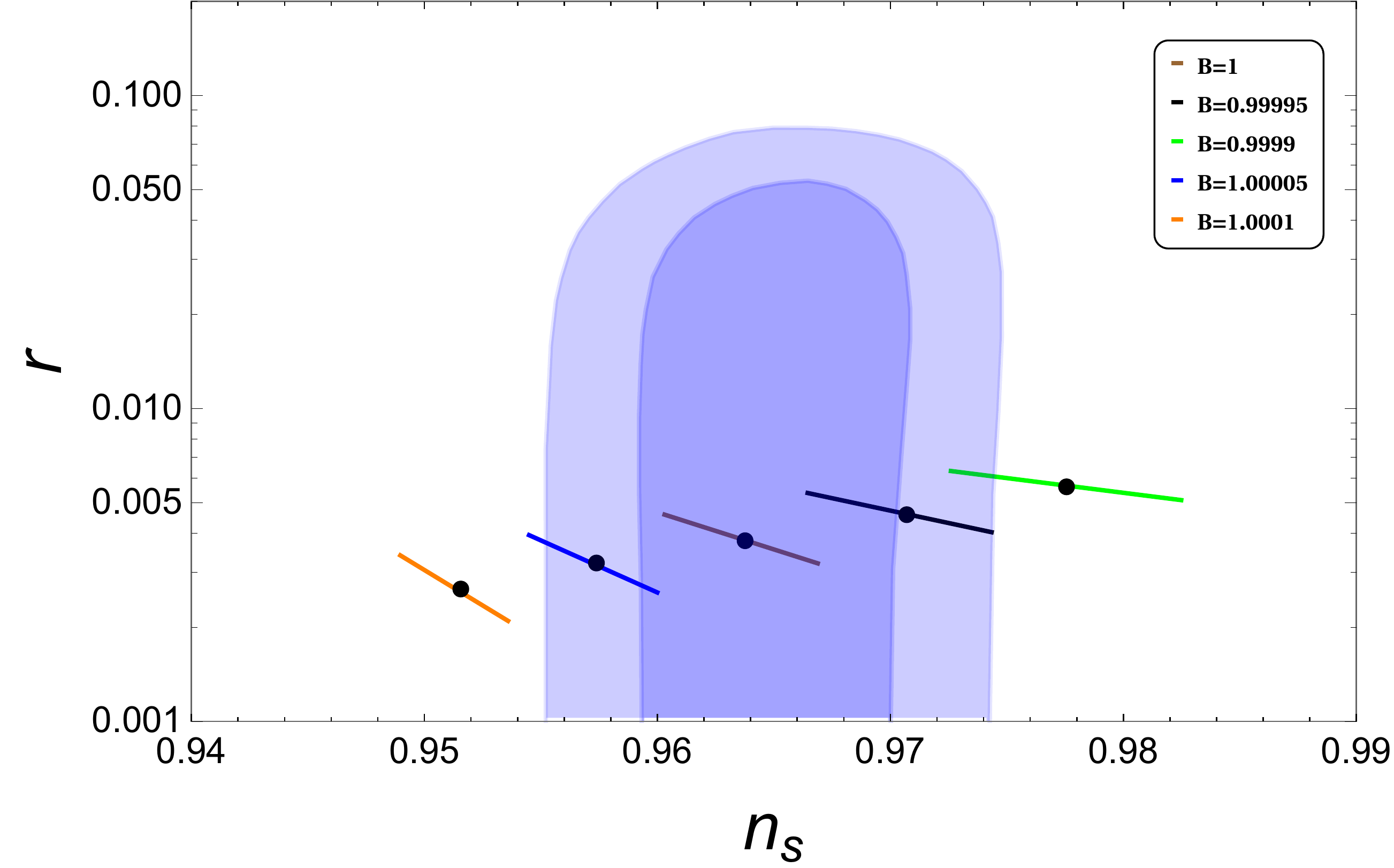}
 \caption{A logarithmic plot (for vertical axis only) for the $n_s - r$ predictions of the inflationary potential through
 Eq. (\ref{Eq:Vinf}). Here scanning over $N_e= 50-60$ for a fixed $B$ value results one segment on top of
 which a black dot is present denoting the prediction for $N_e = 55$ only. The color codes are in accordance with Fig. \ref{MEig}
 representing different values of parameter $B(\xi)$. Planck observation regions, the dark and light blue regions, correspond
 to 1 and 2 $\sigma$ contour of the Planck 2018 (TT, TE, EE + LowE + Lensing + BK14)  data \cite{Akrami:2018odb}.}
 \label{fig:planck1}
\end{figure}
Let us now proceed to determine the inflationary observables in this scenario. The inflationary potential in Eq.(\ref{InfPA}) contains
two free parameters $A$ and $\xi$. Among them, $\xi$ takes part in determining $r$ and $n_s$. The other parameter $A$ will be fixed by observed value of scalar perturbation spectrum $P_s=2.2\times 10^{-9}$. In Fig. \ref{fig:planck1} we show the Logarithmic plot of the spectral index $n_s$  versus the tensor to scalar ratio $r$, as predicted by our model. We also use the Planck limits for comparison purpose. The brown, black and green curves represent the predictions for $n_s$ and $r$ corresponding
to values of $B= 1, 0.99995$ and $0.9999$ respectively. Similarly the blue and orange lines are for $B= 1.00005$ and $1.0001$
respectively. The color codes are in accordance with Fig.\ref{MEig}. In this plot, a single colored line segment represents the variation of
the number of $e$-foldings $N_e$ from 50 to 60, where the prediction for $N_e = 55$ is denoted by a black dot over the
respective line. { Note that due to the presence of another minimum at a large field value for $\chi_1$
in case with $B > 1$ (see Fig. \ref{MEig}), an initial condition on the field value of the inflaton has to be set. For such choices
of $B>1$ we assume that near the onset of inflation, inflaton starts with not so large field value, rather it was close
enough to the flat part ($i.e.$ near maximum) of the potential $\lesssim 8$ (in $M_P$ unit). Then it can slowly roll toward
the minimum at origin and inflation can be realized.}

\begin{table}[h]
\begin{center}
\begin{tabular}{|c |c | c | c | c | c |}
\hline
Sl no. & $A$ & $\xi$ & $x^*$ & $n_s$ & $r$\\
\hline
I & $1.3\times 10^{-10} $ & 0 & $6.13$ & $0.96387$  & $0.0038$ \\
\hline
II &$1.55\times 10^{-10} $ & $5\times 10^{-5}$ & $6.20$ &  $0.97066$
& $0.0046$\\
\hline
III & $1.85\times 10^{-10} $ & $1\times 10^{-4}$ & $6.25$ & $0.97795$
& $0.0056$  \\
\hline
IV & $1.07\times 10^{-10} $ & -$5\times 10^{-5}$ & $6.07$ & $0.95735$
& $0.0032$  \\
\hline
V & $0.9\times 10^{-10} $ & -$1\times 10^{-4}$ & $6.02$ & $0.95156$ &
$0.0026$  \\
\hline
\end{tabular}
\end{center}
\caption{Inflationary predictions ($n_s - r$) for five reference points
in our set up considering $N_e=55$. These points are also
indicated by black dots in Fig. \ref{fig:planck1}.}
\label{tab:infP}
\end{table}
Here we tabulate few reference points which provide correct values of $n_s$ and $r$ within the allowed range of Planck limit considering $N_e=55$. Values of $A$ are fixed from the value of the required power spectrum. Note that the parameters
$A$ and $\xi$ in Table \ref{tab:infP} are simply combinations of the original variables: $c$, $\mu$, $\lambda$ and $M_*$.
All of these variables serve significant importance from the model point of view. Therefore we should also
estimate their magnitude in the set up. For the purpose we consider $M_*=M_P$, argued as the natural
choice in the introduction.

Corresponding to the  reference points (I-V) in Table \ref{tab:infP}, below in Table \ref{tab:OthersV} we provide values
of $c$ and $\mu$ in $M_P=1$ unit for different values of $\lambda$. { Note that $\lambda$ being
the parameter associated with the explicit $R$-symmetry breaking term, it is expected to be small. Hence in obtaining
$c$ and $\mu$ values, we have kept $\lambda\ll 1$.}
\begin{table}[H]
\begin{center}
\begin{tabular}{ |c|c|c|c| }
\hline
Sl. no. & $\lambda$ & $c$& $\mu$ \\ \hline
\multirow{2}{*}{I} & $10^{-4}$ & $4.81481 \times 10^{-4}$& $1.44444\times 10^{-7}$ \\
 & $(10^{-6})$ &$(4.81481)$ &  $(1.44444\times 10^{-5})$ \\ \hline
\multirow{2}{*}{II} & $10^{-4}$ & $5.74017\times 10^{-4}$& $1.72214\times 10^{-7}$ \\
 & $(10^{-6})$ &$(5.74017)$ &  ($1.72214\times 10^{-5}$)\\ \hline
\multirow{2}{*}{III} & $10^{-4}$ & $6.85048\times 10^{-4}$& $2.05535 \times 10^{-7}$ \\
 & $(10^{-6})$ &$(6.85048)$ &  $(2.05535 \times 10^{-5})$ \\ \hline
\multirow{2}{*}{IV} & $10^{-4}$ & $3.96336\times 10^{-4}$& $1.18895\times 10^{-7}$ \\
 & $(10^{-6})$ &$(3.96336)$ &  $(1.18895\times 10^{-5})$ \\ \hline
\multirow{2}{*}{V} & $10^{-4}$ & $3.3340\times 10^{-4}$& $1.0001\times 10^{-7}$ \\
 & $(10^{-6})$ &$(3.3340)$ &  $(1.0001\times 10^{-5})$ \\ \hline
\end{tabular}
\end{center}
\caption{Values of $c$ and $\mu$ (in $M_P$ unit) with $M_*=M_P$  for five reference
points from Table \ref{tab:infP}.}
\label{tab:OthersV}
\end{table}
It can be noted from Table \ref{tab:OthersV} that there exists a correlation between the two mass parameters $c$ and
$\mu$ for different values of $\lambda$. For example, in reference point I of Table \ref{tab:OthersV} with { $\lambda
\sim 10^{-4}$},
the values of $c$ and $\mu$ are found to be {$\sim 4.81\times 10^{-4}$}
 and {$\sim 1.44\times 10^{-7}$} respectively. For a
comparatively smaller value of {$\lambda\sim 10^{-6}$}, magnitudes of $c$ and $\mu$ become {$\mathcal{O} (1)$}
 and { $\mathcal{O} (10^{-5})$} respectively.
 This can be interpreted by looking at the expressions of $A$ and $B$ which involve all the parameters $c, \mu, \lambda,
 M_* = M_P$ and keeping in mind that in order to achieve successful inflation, we have to have $B = 3 \lambda c /\mu$ value
 very close to unity. Therefore with a fixed choice of $\lambda$, the ratio $c /\mu$ is uniquely fixed. Then the parameter
 $A = 3 \mu^2/c$ will fix the value of $\mu$ from the requirement that the power spectrum $P_s \sim \mathcal{O}(10^{-9})$.

{

It is also important to discuss the value of the modulus field vev ($c = \langle T + T^* \rangle$) in terms of high scale
dynamics. In fact the KKLT  scenario assumes \cite{Kachru:2003aw,Balasubramanian:2005zx} that the volume modulus
field can be stabilized by non-perturbative corrections to the superpotential  that arise  from instanton effects or gaugino condensation. Considering a single modulus as in the KKLT model \cite{Kachru:2003aw}, the non-perturbative part of the superpotential $W_{np} \sim e^{-a T}$ where $a$ is a positive constant. It is also emphasized there in \cite{Kachru:2003aw}
that the condition $aT\gg 1$ should be ensured in order to have control over supergravity approximation. Now depending
on the magnitude of $a$, the vev of the modulus field $T$ could be bigger or smaller than one (in $M_P=1$ unit). In case
$\langle T\rangle$ is sub-Planckian, $a$ must be greater than one (to maintain the condition $aT\gg1$). Such a  case is
discussed in ref.  \cite{Lust:1991yi} where it is shown that $a>1$ can be realized through the choice of a hidden sector
gauge group in order to perform the gaugino condensation. On the other hand to establish $\langle T\rangle\gtrsim 1$,
$a$ could be both bigger or smaller than one. In our scenario both the cases regarding the moduli vev ($ c <1$ or $>1$) can be accommodated as displayed in Table \ref{tab:OthersV} depending on the magnitude of $\lambda$.
For example, $\lambda\sim 10^{-4}$ can make the modulus vev $c\sim \mathcal{O}(10^{-4})$, while for $\lambda\sim
10^{-6}$, $c$ could be $\mathcal{O}(1)$ or more.}

For all the reference points mentioned in Table  \ref{tab:infP}, numerically it is found that the mass of the $\chi_2$ field
during inflation is significantly higher compared to the Hubble scale ($H$) during inflation given by  $H^2 = V_{\textrm{Inf}}/3$. In particular, one finds $m^2_{\chi_2}/H^2 \simeq 4$ at the minimum of $\chi_2$
($\chi_2 = 0$). Hence $\chi_2$ would be stabilized at origin during inflation. Furthermore we have also found numerically
that the slope along the $\chi_2$ direction is much steeper compared to the one for $\chi_1$. Hence $\chi_2$ will move
faster and reaches the minimum much earlier than $\chi_1$. This justifies our assumption $\chi_2 = 0$ during inflation. Inflaton
mass ($m_{\chi_1}$) at its minimum for the above mentioned points is $\mathcal{O}(10^{13})$ GeV as expected for this
type of inflation scenario. We end this section by observing that even if $\chi_1$ is super-Planckian as required by the
slow-roll condition, the field $S$ remains sub-Planckian as seen from Eq. (\ref{refield}). Now $S$ being sub-Planckian,
higher order $U(1)_R$ breaking terms in $W$ are accordingly less important.

\section{Conclusion}

In this paper, we propose a global $R$ symmetry motivated inflation model within no-scale SUGRA.
We find that the minimal $U(1)_R$ symmetric superpotential (quadratic in inflaton superfield) is unable to provide
a successful inflation as the associated scalar potential turns out to be extremely steep. Then we introduce an
explicit $R$ symmetry breaking term in the superpotential at a non-renormalizable level which provides the required
flatness for inflation. The introduction of such a $U(1)_R$ breaking term is motivated by the fact the any global
symmetry will be broken by the gravity effect. For this reason, we associate the cut-off scale of this non-renormalizable
term with $M_P$. The effective inflation potential resulted from our proposed
set-up carries similarity with  Staborinsky like inflation models in the limit, one combination of parameters of the superpotential
and no-scale Kahler potential as $B = 1$. Varying $B$ from unity by tiny amount leads to the predictions for the
spectral index and tensor-to-scalar ratio. In order to keep these predictions within the limit allowed by Planck data,
we evaluate the magnitudes of the relevant mass parameters of the model. Such a construction involving explicit
$R$ symmetry breaking term may also have some interesting consequences while supersymmetry breaking will also
be involved.  Since any dynamical supersymmetry breaking model requires that $R$ symmetry should spontaneously
be broken leading to the presence of $R$ axion, such an explicit breaking term, connected with inflation, in our set-up
can be helpful in providing the mass of it.

\section*{Acknowledgments}

The work of S.K. and A. M. is partially supported by the STDF project 18448 and the European Union FP7 ITN INVISIBLES (Marie Curie Actions, PITN-GA-2011-289442).


\end{document}